\documentclass[twocolumn,twoside,slac_two]{revtex4}
\usepackage{graphicx}
\usepackage{fancyhdr}
\newcommand{\im}{\ensuremath{\overline{m}}}

\begin{document}

\title{Radio Variability Studies of Gamma-Ray Blazars with the OVRO 40~m Telescope}

%

\author{
  J.~L. Richards}\altaffiliation{Current address: Department of Physics, Purdue University, 525 Northwestern Ave, West Lafayette, IN 47907, USA}
\author{W. Max-Moerbeck}
\author{V. Pavlidou}
\author{A.~C.~S. Readhead}
\author{T.~J. Pearson}
\author{O.~G. King}
\author{R. Reeves}
\author{M.~A. Stevenson}
\author{M.~C. Shepherd}
\affiliation{Cahill Center for Astronomy and Astrophysics, California Institute of Technology, 1200 E. California Blvd., Pasadena, CA 91125, USA}

\begin{abstract}
  Since late 2007, we have been monitoring a large sample of known and
  likely gamma-ray--loud blazars at 15~GHz twice per week with the
  Owens Valley Radio Observatory (OVRO) 40~m Telescope.  Our initial
  sample included the 1158 sources above declination $-20^{\circ}$
  from the Candidate Gamma-Ray Blazar Survey (CGRaBS), and we have
  since added nearly 400 more sources, including all blazars
  associated with \emph{Fermi} Large Area Telescope (LAT) detections
  in the First AGN Catalog (1LAC).  Here, we describe the new sample
  and present results for 2008 through early 2011.  Using statistical
  likelihood analyses, we compare the variability amplitude for
  various sub-populations within our sample. These include comparisons
  of gamma-ray--loud versus gamma-ray--quiet objects, BL~Lac objects
  versus flat-spectrum radio quasars, and a study of the variability
  amplitude trend with redshift.  We also describe KuPol, the new
  digital Ku-band receiver being constructed for the 40~m
  telescope. This new receiver will provide total intensity and linear
  polarization measurements over the 12--18~GHz band, with 16~MHz
  spectral resolution.

\end{abstract}

\maketitle

\thispagestyle{fancy}


\section{INTRODUCTION}

Blazars are the most extreme class of active galactic nuclei (AGN),
probably resulting when an AGN is viewed along its relativistic jet
axis. Although a variety of models have been proposed, and despite
decades of observation, many fundamental questions about blazars and
AGN physics remain open.  Among them:
\begin{itemize}
\item How are jets launched, accelerated, collimated, and confined?
\item Of what are the jets composed?
\item Where in the jet are the observed photons emitted?
\item What are the details of the emission mechanisms?
\end{itemize}
Gamma-rays are our best tool for probing the most extreme processes
within AGN, and coordinated multiwavelength observations are crucial
to the interpretation of gamma-ray data.

\subsection{OVRO 40~m Program}
Since late 2007, we have undertaken a continuous, fast-cadence (twice
weekly per source) monitoring program beginning with the
systematically-selected sample of 1158 blazars north of
$\delta=-20^{\circ}$ from the Candidate Gamma-Ray Blazar Survey
(CGRaBS)~\cite{healey_cgrabs:all-sky_2008}.  The sample has grown to
more than 1550 sources, including all 454 ``clean'' AGN associations
from the first-year \emph{Fermi} Large Area Telescope (LAT) AGN catalog
(1LAC)~\cite{abdo_first_2010}. Newly LAT-detected AGN are
added to the sample as well. Results for the CGRaBS sample during the
2008--2009 period are presented and analyzed and the program is
described in detail in \cite{richards_blazars_2011}. Radio light
curves for the CGRaBS sample are available to the public for download from {\url{http://www.astro.caltech.edu/ovroblazars}}.

\section{RADIO VARIABILITY}

To provide a robust measure of the amplitude of variability observed
in a light curve, we introduced the \emph{intrinsic modulation index},
$\im\equiv\sigma_0/S_0$, where $\sigma_0$ and $S_0$ are the standard
deviation and mean that would be measured for a light curve using a
noiseless receiver with perfect sampling. In
\cite{richards_blazars_2011} we derive a likelihood method for
estimating \im{} and its uncertainty from an observed light
curve. Unlike the various measures of variability amplitude commonly
used in the literature, the intrinsic modulation index provides a
rigorous measure of its uncertainty which is essential for statistical
comparisons.

We computed \im{} for each CGRaBS source using two years of data in
\cite{richards_blazars_2011}. Using 3.2 years of data, we find that
the intrinsic modulation index for each source is typically either
consistent within uncertainties or is found to be larger with the
longer data set. An examination of the sources with the most
significant changes shows that the significant increases typically
occur in sources that demonstrated no or little variability during the
first two years and subsequently exhibited a major flare or change in
brightness. This suggests that the two-year data set was simply not
long enough to capture the full range of variability behavior of the
sources. Previous findings indicate that 4--6~year or longer
timescales for flaring activity are common in
blazars~\cite[e.g.,][]{hovatta_statistical_2007}, so further increases
in the modulation indices are likely as the monitoring program
continues.

\subsection{Population Studies}

\begin{figure}[tb]
  \centering
  \includegraphics[width=0.8\columnwidth]{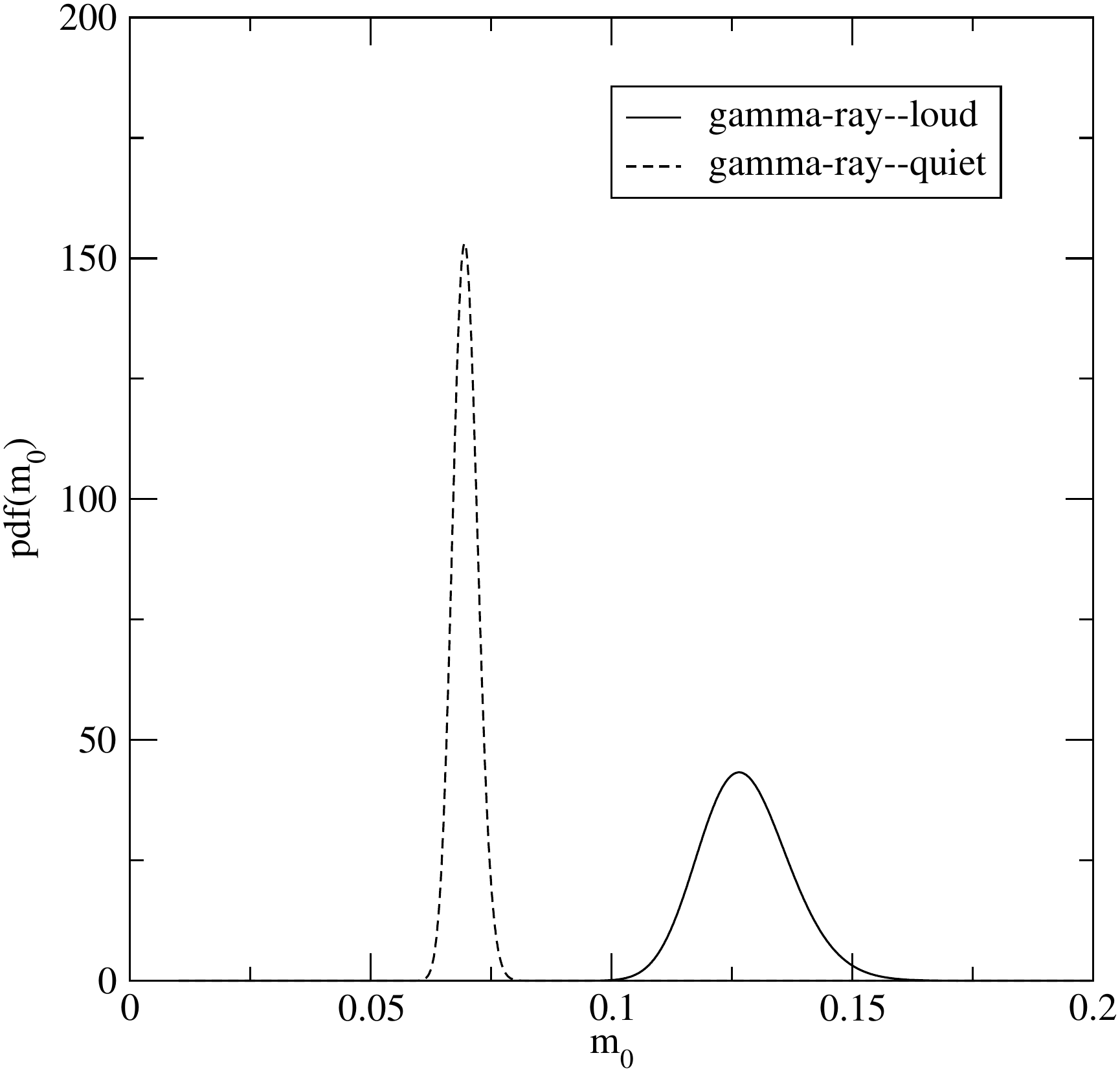}
  \caption{Likelihood distributions for the two-year mean intrinsic
    modulation index, $m_0$, for the CGRaBS in 1LAC (solid) and not in
    1LAC (dashed). The gamma-ray loud 1LAC subset are 5.7\% more
    variable on average, a $> 6\sigma$ significant difference.}
  \label{fig:gamma_nongamma} 
\end{figure}

In \cite{richards_blazars_2011}, we found strong ($> 6\sigma$)
evidence that the gamma-ray--loud subset of CGRaBS (i.e., those
sources that appeared in the 1LAC catalog) was, on average, more
variable at 15~GHz than was the gamma-ray--quiet subset (see
Figure~\ref{fig:gamma_nongamma}). This trend persists in the longer
data set with similar significance. This establishes the first
rigorous demonstration of a connection between gamma-ray emission and
radio variability amplitudes in blazars.

\begin{figure}[tb]
  \centering
  \includegraphics[width=0.8\columnwidth]{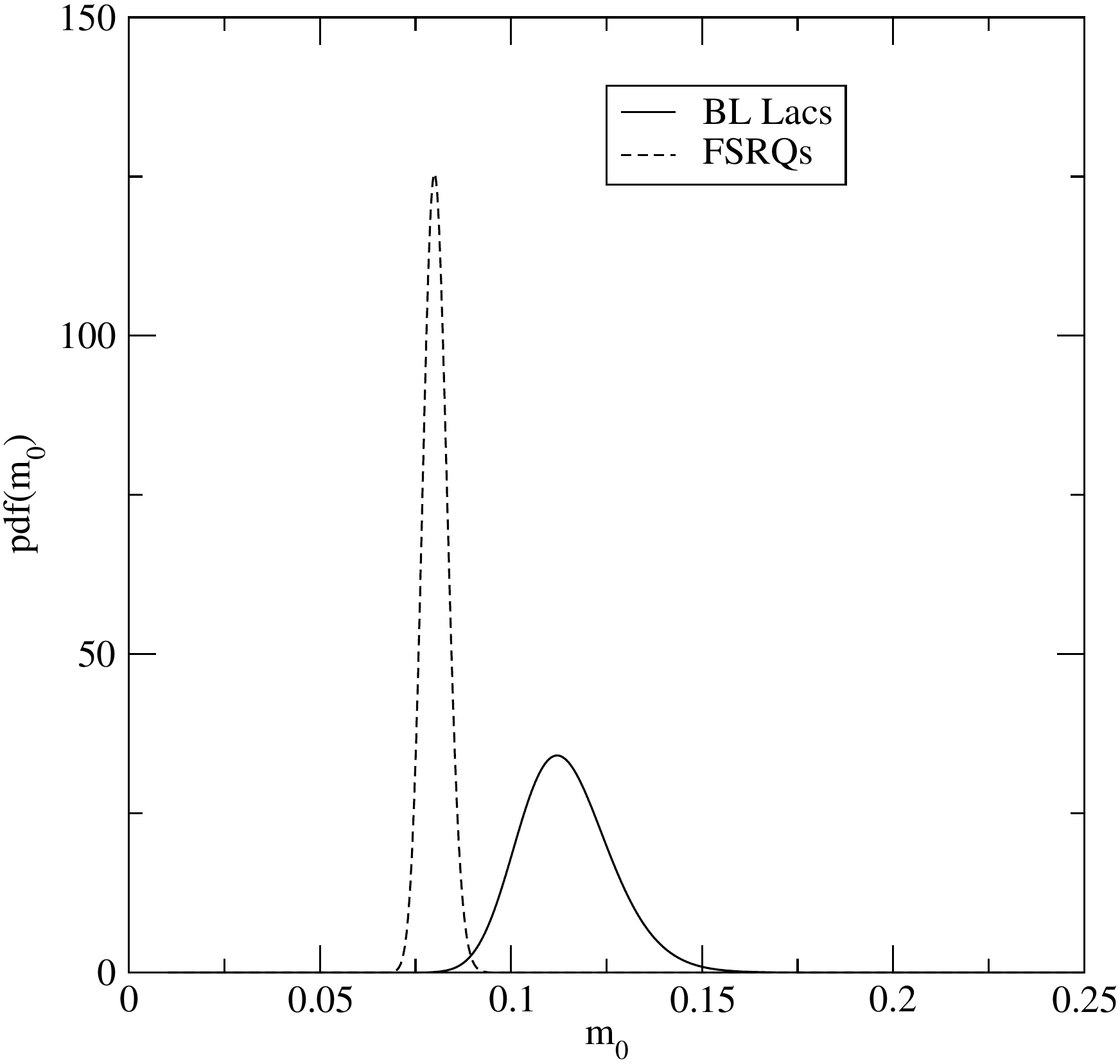}
  \caption{Likelihood distributions for the two-year mean intrinsic
    modulation index, $m_0$ for the CGRaBS BL~Lac object (solid) and
    FSRQ (dashed) subsamples. The BL~Lac objects are 3.2\% more
    variable on average, a $>3\sigma$ significant difference.}
  \label{fig:bll_vs_fsrq}
\end{figure}

\begin{figure}[tb]
  \centering
  \includegraphics[width=0.8\columnwidth]{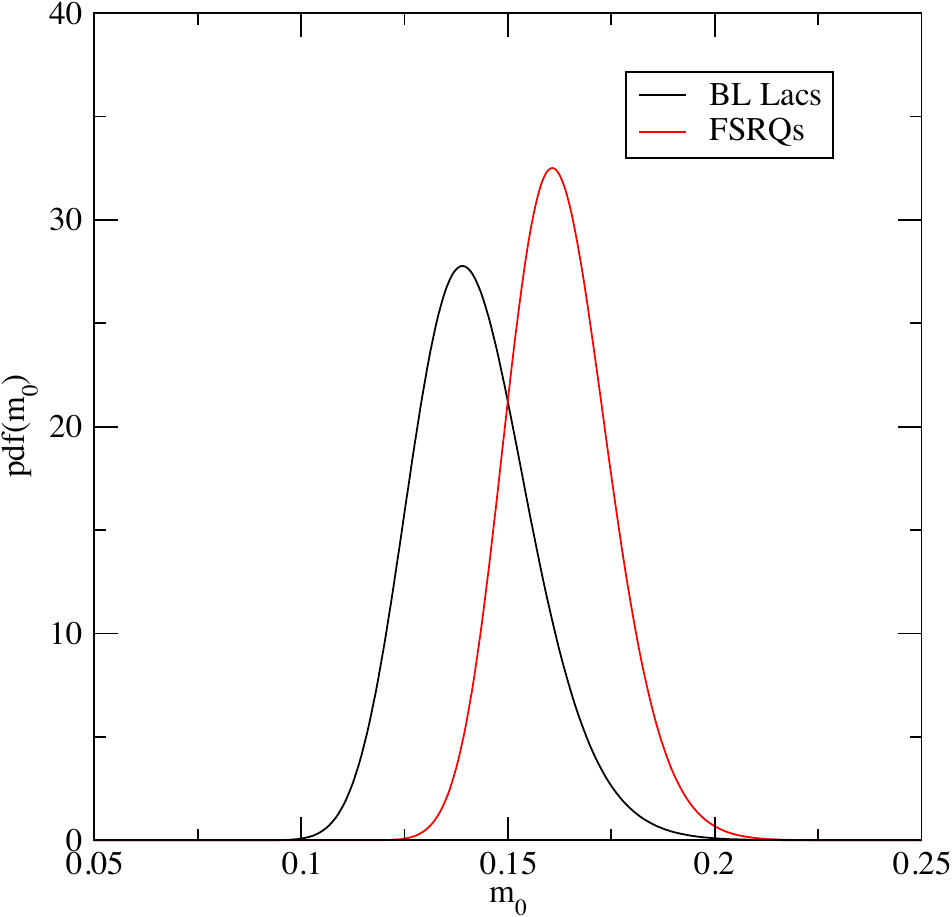}
  \caption{Likelihood distributions for the two-year mean intrinsic
    modulation index, $m_0$ for the 1LAC BL~Lac object (black) and
    FSRQ (red) subsamples. The two distributions are consistent with
    equal variability at the $1.5\sigma$ level.}
  \label{fig:bll_vs_fsrq_1lac}
\end{figure}

In Figure~\ref{fig:bll_vs_fsrq}, we compare the two-year variability
amplitudes for CGRaBS sources identified as BL~Lac objects with those
identified as flat-spectrum radio quasars (FSRQs). The BL~Lac objects
are found to vary more strongly than the FSRQs with $>3\sigma$
significance. A similar result is found with the longer data set for
CGRaBS sources. However, for the 1LAC sample, using 3.2~years of data
we find the BL~Lac objects and FSRQs to be consistent with the same
variability amplitude within $\sim1.5\sigma$ (see
Figure~\ref{fig:bll_vs_fsrq_1lac}). Further study of this effect in
\cite{richards_thesis_2012} suggests that the gamma-ray--loud FSRQs
are much more radio variable than gamma-ray--quiet FSRQs whereas
BL~Lac objects display similar radio variability whether or not
gamma-ray--loud.

\subsection{Redshift Trend}

\begin{figure}[tb]
  \centering
  \includegraphics[width=0.8\columnwidth]{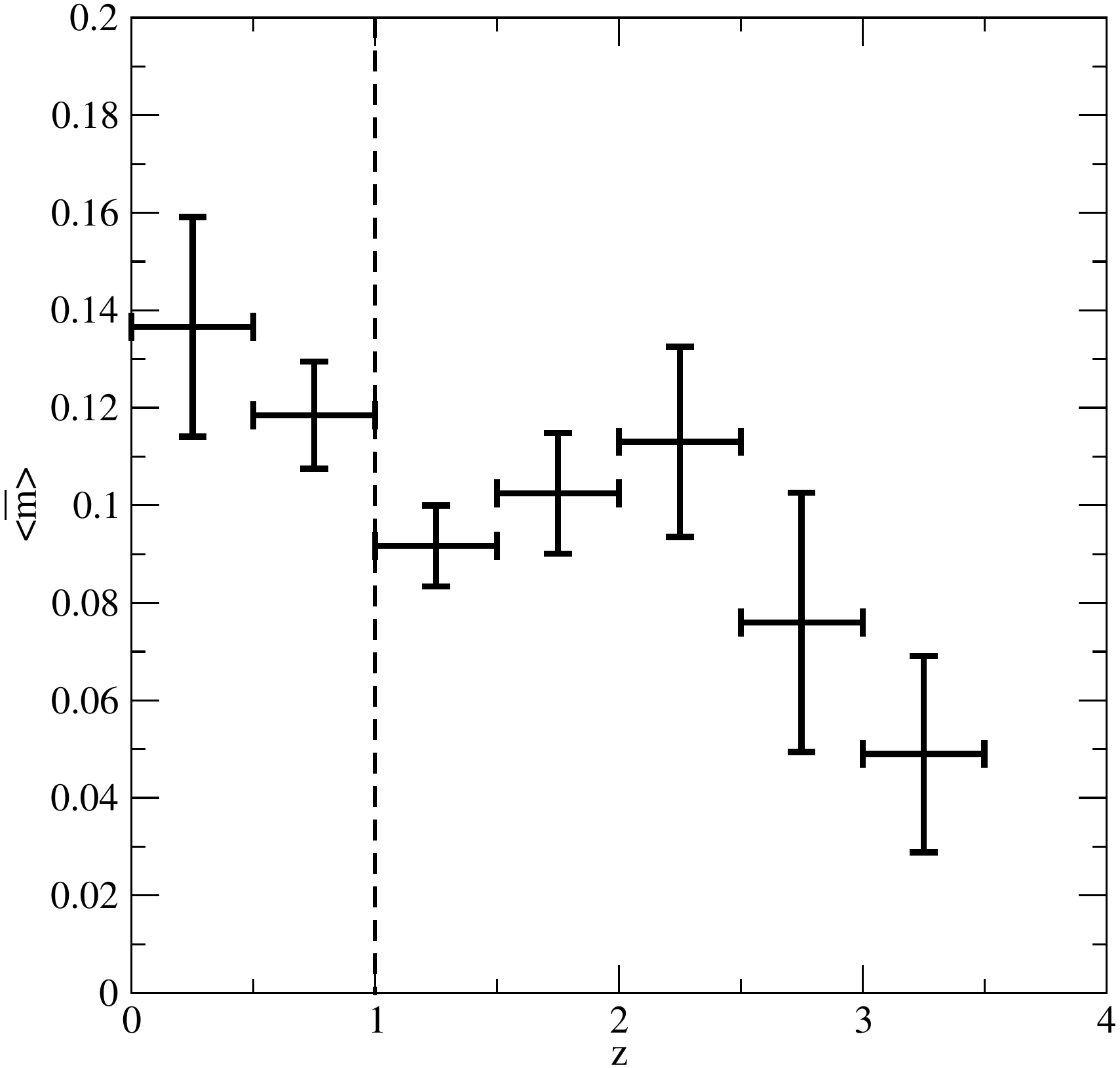}
  \caption{Average intrinsic modulation indices for CGRaBS FSRQs
    separated in redshift bins of width $\Delta z=0.5$. Error bars
    indicate the rms scatter within each bin.}
  \label{fig:m_vs_z}
\end{figure}

To examine the relationship between radio variability amplitude and
redshift, we restrict our sample to the FSRQs. This is necessary
because the lack of optical lines in BL~Lac objects makes redshift
determination difficult, resulting in $\sim50\%$ or less redshift
completeness in our sample. This also prevents the difference in
variability already found between BL~Lac objects and FSRQs from
contaminating our comparison.  In \cite{richards_blazars_2011}, we
found an apparent trend of decreasing variability amplitude with
increasing redshift, shown in Figure~\ref{fig:m_vs_z}.  A comparison
of bright ($S_0>400~\mathrm{mJy}$) FSRQs at $z<1$ with those at $z>1$
found the lower-redshift FSRQs to be more variable with $3\sigma$
significance~\cite{richards_blazars_2011}.

\begin{figure}[tb]
  \centering
  \includegraphics[width=0.8\columnwidth]{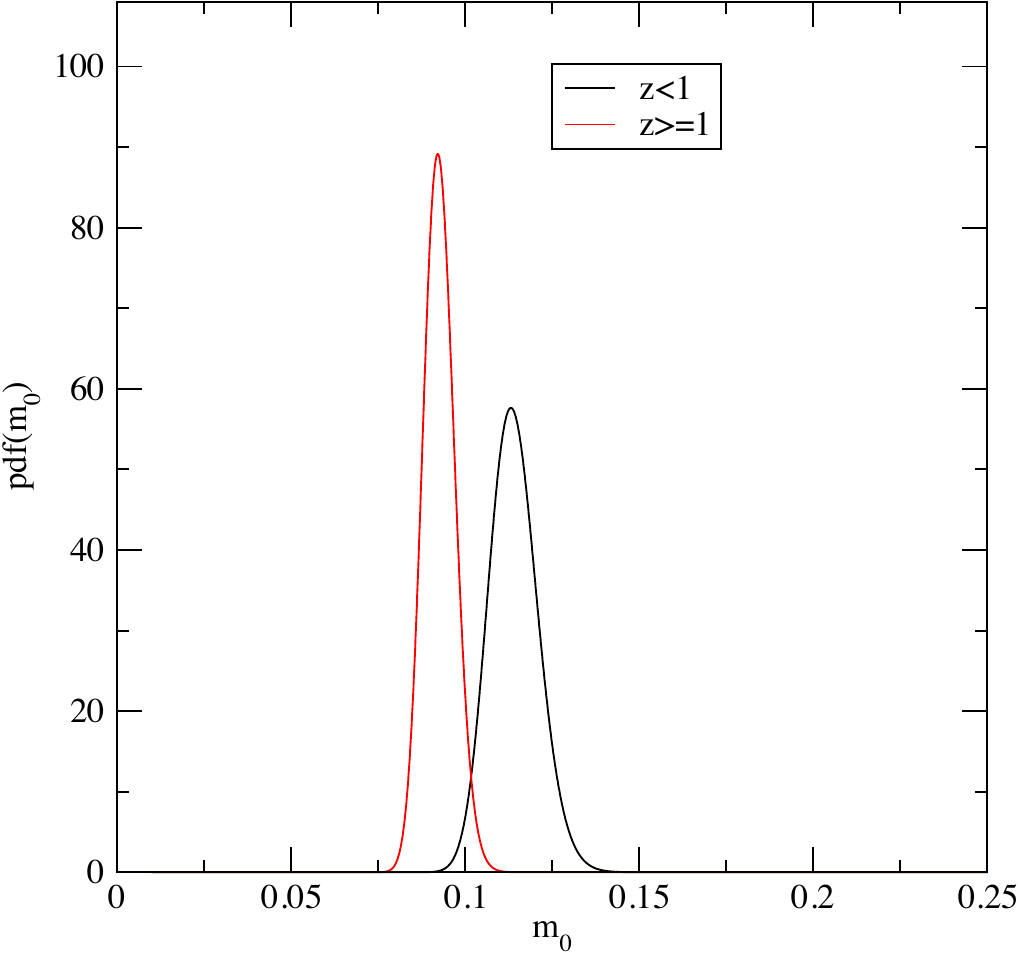}
  \caption{Likelihood distributions for the intrinsic modulation index
    for low ($z<1$, black) and high ($z>1$, red) redshift subsets of
    the bright ($S_0>400~\mathrm{mJy}$) 1LAC FSRQ sample. The apparent
    difference is not statistically significant.}
  \label{fig:z_high_low_1lac}
\end{figure}

While this trend continues with the 3.2~year CGRaBS data, the
significance of the difference is less with the additional data. We
also find that FSRQs within the 1LAC sample do not show a significant
difference in variability between the high and low redshift subsamples
(Figure~\ref{fig:z_high_low_1lac}).  In \cite{richards_thesis_2012},
we find that this significance continues to decrease with additional
data, suggesting that this observed difference is likely spurious.  If
real, however, a number of competing effects must be removed to
determine whether the observed trend reflects any intrinsic
variability trend with redshift.

\subsection{Cross-Correlation}

A major aim of the OVRO monitoring program is the detailed
cross-correlation of radio light curves with gamma-ray light curves
from the LAT. Peaks in the cross-correlation function could
identify the relative locations of gamma-ray and radio emission within
a blazar jet.  However, establishing the \emph{physical} significance
of an apparent correlation is a major challenge. Typical methods for
computing the significance of a cross-correlation peak do not account
for correlation between points within each light curve and, as a
result, seriously overestimate the significance of an apparent
peak. We have developed a Monte Carlo method to evaluate the
significance of apparent correlations more accurately. This method and
early results are described in more detail
in~\cite{max-moerbeck_relation_2010} and in Max-Moerbeck et al. in
these proceedings. A detailed study of cross-correlation between OVRO
15~GHz light curves and LAT gamma-ray light curves is
forthcoming~\cite{max-moerbeck_forthcoming}.

\section{KUPOL RECEIVER UPGRADE}

Blazars characteristically exhibit polarized radio emission which can
reveal the configuration of the magnetic field in the synchrotron
emission regions. As a result, monitoring radio polarization in
addition to intensity can add valuable information and provide an
additional channel to study the relationship between emission in the
radio and other bands. To this end, we have designed and begun
construction of a new Ku-band (15~GHz) receiver, ``KuPol,'' with both
linear polarization and total intensity measurement
capabilities. Using a correlation receiver architecture and digital
back-end, KuPol provides 16~MHz spectral resolution over a total
bandwidth from 12--18~GHz using the same beam-differenced optical
design as the current program. Construction of KuPol is nearly
complete and commissioning is planned for early 2012.

\section{CONCLUSIONS}

We are actively and continuously monitoring a sample of more than 1550
blazars, including all \emph{Fermi} gamma-ray blazars north of
$\delta=-20^{\circ}$ twice per week at 15~GHz. We have rigorously
demonstrated a connection between gamma-ray emission and the strength
of radio variability in our sample, an important prerequisite for
establishing the physical significance of correlations between radio
and gamma-ray light curves. In addition, we have developed and begun
applying a Monte Carlo method for accurately evaluating the
statistical significance of apparent peaks in the discrete
cross-correlation function between radio and gamma-ray light
curves. In addition to continuing our monitoring program, we will soon
deploy KuPol, adding an increased bandwidth, spectral resolution, and
linear polarization monitoring to our radio monitoring program. These
tools and the ever growing data sets will enable detailed study of the
connection between radio and gamma-ray emission in blazars.

\bigskip 
\begin{acknowledgments}
  This work is supported in part by NASA grants NNX08AW31G and
  NNG06GG1G and NSF grant AST-0808050.
\end{acknowledgments}

\bigskip 

\begin{thebibliography}{9}   

\bibitem{abdo_first_2010} Abdo, A.~A. et al. 2010, ApJ, 715, 429.
\bibitem{healey_cgrabs:all-sky_2008} Healey, S.~E. et al. 2008, ApJS, 175, 97.
\bibitem{hovatta_statistical_2007} Hovatta, T. et al. 2007, A\&A, 469, 899.
\bibitem{max-moerbeck_relation_2010} Max-Moerbeck, W. et al. 2010, ``The Relation Between the Radio and Gamma-Ray Emission in Blazars from 15~GHz Monitoring with the OVRO 40~m Telescope and Fermi-GST Observations,'' in Proceedings of the Workshop ``Fermi Meets Jansky: AGN in Radio and Gamma-Rays,'' Bonn (Germany), 21--23 June 2010, p. 77.
\bibitem{max-moerbeck_forthcoming} Max-Moerbeck, W. et al., 2011,
  \emph{in preparation}.
\bibitem{richards_blazars_2011} Richards, J.~L. et al. 2011, ApJS, 194, 29.
\bibitem{richards_thesis_2012} Richards, J.~L. 2012, Ph.D Thesis,
  California Institute of Technology.


\end{thebibliography}

\end{document}